\documentclass[journal abbreviation]{copernicus}

\begin{document}

\title{Is current disruption associated with an inverse cascade?}

\author[1]{Z. V\"or\"os}
\author[2]{A. Runov }
\author[1]{M.P. Leubner}
\author[3]{W. Baumjohann}
\author[3]{M. Volwerk}

\affil[1]{Institute of Astro- and Particle Physics, University of Innsbruck, Innsbruck, Austria.}
\affil[2]{UCLA, Los Angeles, USA.}
\affil[3]{Space Research Institute, Austrian Academy of Sciences, Graz, Austria.}

\runningtitle{An inverse cascade?}

\runningauthor{V\"or\"os}

\correspondence{Z. V\"or\"os\\ (zoltan.voeroes@uibk.ac.at)}

\received{}
\pubdiscuss{} 
\revised{}
\accepted{}
\published{}


\firstpage{1}

\maketitle

\begin{abstract}
Current disruption (CD) and the related kinetic instabilities in the near-Earth magnetosphere represent
physical mechanisms which can trigger multi-scale substorm activity including global reorganizations
of the magnetosphere. \citet{lui08} proposed a CD scenario in which the kinetic scale linear modes
grow and reach the typical dipolarization scales through an inverse cascade. The experimental verification
of the inverse nonlinear cascade is based on wavelet analysis.
In this paper the Hilbert-Huang transform is used which is suitable
for nonlinear systems and allows to reconstruct the time-frequency
representation of empirical decomposed modes in an adaptive manner.
It was found that, in the \citet{lui08} event, the modes evolve
globally from high-frequencies to low-frequencies. However, there
are also local frequency evolution trends oriented towards
high-frequencies, indicating that the underlying processes involve
multi-scale physics and non-stationary fluctuations for which the
simple inverse cascade scenario is not correct.
\end{abstract}


\introduction
A key element of substorm physics is an enhanced transport of magnetic flux from
the dayside  magnetopause over  the poles  to the  magnetotail during the growth
phase  of  substorms \citep{baum99}. In  the  tail  the  magnetic  energy  is  accumulated and
subsequently abruptly released during the expansion phase. In general, substorms
are associated  with multi-scale  processes including  electron- (tens  of kms),
ion-  (hundreds  of kms),  MHD-  ($\geq$ thousands  of  kms) scales  and  global
reconfigurations of the  magnetosphere \citep{naka06, lait07}. For this  reason, the identification  of
location(s)  and  timing  of  substorm  triggering  mechanisms  was  not   fully
conclusive. Magnetic reconnection (MR), converting magnetic energy into  kinetic
energy of accelerated plasma  and energetic particles at  20 to 30 $R_E$,  could
explain many substorm signatures, such  as magnetic field dipolarization due  to
flow breaking at the  barrier of strong dipolar  field (at $\sim 10  R_E$), flux
pile-up  and  the tailward  motion  of dipolarization  front,  the formation  of
substorm current wedge, and the associated auroral and ground based activities \citep{baum99}.
On the other hand, the substorm triggering scenario called current disruption  (CD)
invokes  kinetic  instability  processes  leading  to  cross-tail  CD   (current
reduction), dipolarization, formation of the substrom current wedge in the  near
-Earth  region at  $\sim 10  R_E$ and  other auroral  and ground-based  substorm
signatures \citep{lui04}.  The  CD  scenario  offers   a  possibility  to  explain  both   the
dipolarization  and  the  tailward  movement  of  dipolarization  front  without
invoking reconnection associated flow breakings  or flux pile-up .
In fact,  some
dipolarizations were observed without  local Earthward plasma flows \citep{lui08}.
Though the
multi-spacecraft timing observations from THEMIS spacecraft indicate predominant
substorm triggering due  to MR \citep{ange08}, both  mid-tail (MR) and  near-Earth (CD) processes
can  contribute to  the global  substorm activity.  For example,  MR associated
Earthward flows might trigger near-Earth CD \citep{runo08}.

CD is associated with kinetic instabilities in rather localized
regions. It  was shown by \citet{lui04} that  a local kinetic theory
of instabilities can account  for the excited waves exhibiting large
growth for long enough  time even within a  thin current  sheet. The
stability   analysis  of   cross-field  ion-drift  driven
instabilities based  on a  two-fluid approach  also showed  that
waves initially excited near the ion cyclotron frequency ($\sim$0.1
Hz, 10 s) can grow and reach the typical  time scales  associated
with  dipolarization ($\sim$0.006-0.008 Hz, 120-170  s). It  was
suggested \citep{lui08}  that the  corresponding experimental
signature would be a developing inverse cascade in wavelet
time-frequency  representation, i.e.,  the evolution  of wave  modes
from  higher (shorter)  to lower  (longer) frequencies  (time
scales) in time. Conversely,  forward  cascades in  turbulence
transfer  energy from large scales to small scales. The dynamics,
driving sources and dissipation of turbulent fluctuations were
reviewed by \citet{boro03}. In the Earth's plasma  sheet turbulence
was  also observed within  reconnection associated bursty flows
\citep{voro04}. However, developing turbulence was not identified,
possibly  because of the  short duration  of events  or due  to the
observation of  already fully developed turbulence with broad-band
fluctuations (with  no time  evolution of characteristic
frequencies). Observations of a developing turbulence would help us
to identify the spatial or temporal scales of turbulent drivers (the
scales of energy input) and to recognize the direction of turbulent
cascades (forward or inverse). Anyhow, this would require to observe
the associated fluctuations over the characteristic scales of
turbulent drivers first, followed by the development of multi-scale
fluctuations over the inertial range of scales.

\section{Limitations of the wavelet analysis}
While the experimental identification of inverse cascades would be
crucial for understanding the real growth rate of excited kinetic
instability modes, the detection based on wavelet analysis has to be
interpreted rather carefully. The main reason is that the widely
used basic wavelet analysis is non-adaptive \citep{hua98}. Once the
basic wavelet is chosen it is used for the whole data set. This is
not a problem when observations with gradual frequency changes,
chirp signals, or wave trains are analyzed. Actually, the harmonic
signals used to test the wavelet identification of inverse cascade
features by \citet{lui08} belong to this category of well treatable
signals. However, sudden jumps (non-stationarity) or nonlinearity
can generate spurious time-frequency-energy distributions in a
wavelet representation. The basic wavelet transform is also linear
\citep{hua98}. Nevertheless, wavelets can be used successfully for
the multi-scale analysis of nonlinear systems \citep{kij06}. The
results of wavelet analysis, however, heavily depend on the mother
wavelet, proper resolution, scale discretization \citep{kij06},
resulting in non-adaptiveness for multi-component non-stationary
data. The effect of non-stationarity is demonstrated in Figure 1a,
b. The test data consisting of a single sine wave with period 1 s
and a sudden jump (Figure 1a) is analyzed using the Morlet wavelet.
The time-wavelet period-energy distribution (wavelet scalogram) in
Figure 1b is showing that the time localization of the sine wave
(between 4 and 5s) is good only at wavelet periods near $\sim$ 0.125
s. The time localization is rather poor at the wavelet period of 1s,
which is the actual sine wave period. Using a different, optimized
wavelet might lead to a slightly better localization of the sine
wave. Certainly, a mother wavelet cannot be equally good for an
unknown combination of periodic or stochastic components in a time
series.
 \begin{figure}
\noindent\includegraphics[width=8.3cm]{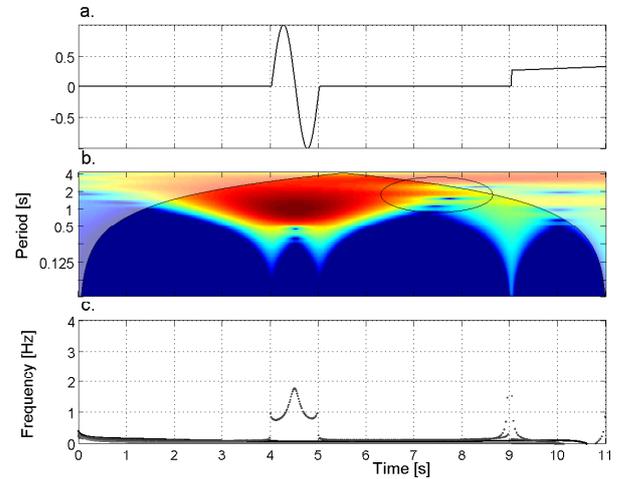}
 \caption{a. Test signal: sine wave + sudden jump; b. Wavelet analysis of the test signal; c. Hilbert-Huang analysis
 of the test signal.}
 \end{figure}

 \begin{figure}
\noindent\includegraphics[width=8.3cm]{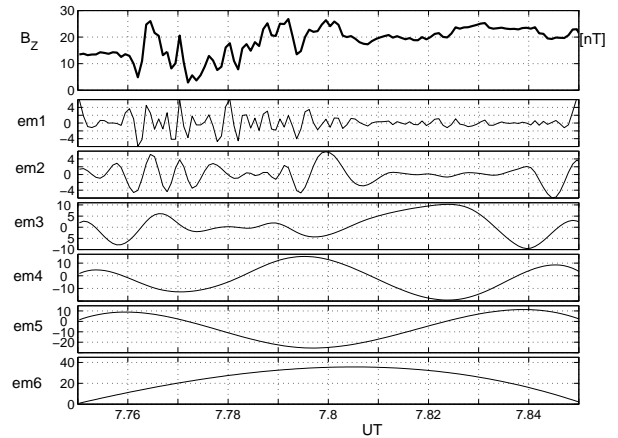}
 \caption{Top subplot: magnetic field $B_Z$ component; Bottom subplots: empirical modes; $emi$s, for $i=1, 2,...6$.}
 \end{figure}

Sudden jumps lead to broad-band energy distributions (Figure 1b).
Since for higher frequencies the basic wavelet is more localized, the sudden jump at
Time=9s is well localized in time over small wavelet period ranges. The time localization worsens
towards longer periods where the basic wavelet is more stretched. The non-stationarity also induces
spurious periodicities in the time-period plane, in our case between Time=7-8 s, where the test signal
contains no fluctuations at all.
Actually, it can be even worse than that. The wavelet identification of scales (periods) can be
entirely misleading. For example,
\citet{hua98} argue that a change occurring over large scales locally in time will appear in
the corresponding time-period distribution over small scales, due to the basic wavelet resolution.
That is, a localized large-scale change (non-stationarity) manifests itself over small scales
in the wavelet spectrum.

\section{The Hilbert-Huang method}
To be able to identify inverse cascade features we need a method which is adaptive
and suitable for non-stationary and nonlinear analysis of multi-scale data.
The Hilbert transform (HT) represents a convolution of time series $X(t)$ with time $1/t$, therefore
accentuates the local properties of $X(t)$. HT is defined as \citep{hua98}

\begin{equation}
Y(t)=\frac{1}{\pi}\textsl{P}\int_{-\infty}^{\infty}\frac{X(\tau)}{t-\tau}d\tau
\end{equation}

where $\textsl{P}$ is the Cauchy principal value of the integral. $X(t)$ and $Y(t)$ are orthogonal and
form a complex conjugate pair

\begin{equation}
Z(t)=X(t)+i Y(t) = a(t)e^{i\Theta (t)}
\end{equation}

$a(t)$ and $\Theta(t)$ are the instantenous amplitude and phase, respectively. The instantenous
angular frequency $\omega(t)$ is the time derivative of $\Theta(t)$.
HT is known for a long time, but the practical value of instantenous frequency $\omega(t)$
was questioned, because of non-unique local values. Local uniqueness can be achieved by putting some
constraints on the data 
(Cohen, 1995). \citet{hua98} proposed a practical decomposition
algorithm, called empirical mode decomposition (EMD), which removes
the ambiguity in determination of $\omega$. The extracted empirical
modes ($em$s) fulfill certain conditions without any loss of
essential nonlinear features of the original data and also ensure
adaptiveness to local signal characteristics \citep{hua98}. The EMD
procedure fits splines to local maxima and minima of $X(t)$, finds
the mean $X_m$ from the two spline functions and computes the
residuals $X_r(t)=X(t)-X_m(t)$. These steps are iterated until a
threshold is reached. The first mode is given by $em1(t) = X_m(t)$.
The next mode is found in the same way as $em1$, but replacing the
original data $X(t)$ with $X(t)-em1(t)$. Successive $emi$ modes are
found by repeating the whole process, until the last residue
contains only a simple trend. Even complex signals can be decomposed
to a finite number of $emi$s. For example, HT based on the EMD was
successfully applied to intermittent turbulence \citep{hu08} just
recently. The extracted $em$ modes represents narrow band stationary
Gaussian processes. Moreover, the modes are locally symmetric
relative to the envelopes defined by the local maxima and minima.
This means that the local means are not computed over a time window
or a local time scale, but are obtained directly from the local
values of the envelopes.
 \begin{figure}
\noindent\includegraphics[width=8.3cm]{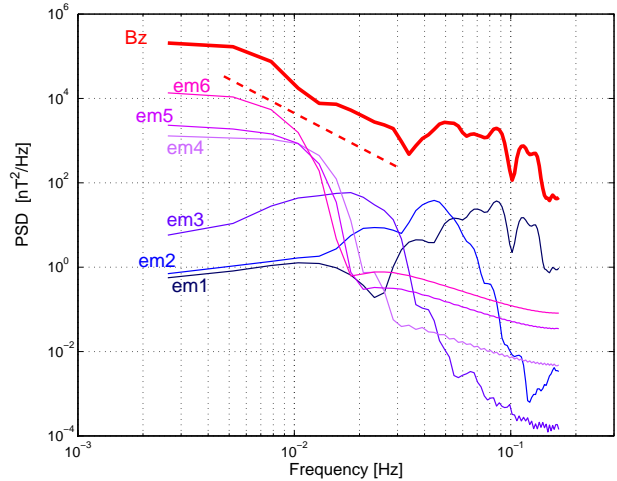}
 \caption{Power spectral density for magnetic field (red thick line)
 and empirical modes; the dashed red line shows the scaling region
 over the inverse cascade frequency range.
}
 \end{figure}

 \begin{figure*}[t]
\begin{center}
\noindent\includegraphics[width=16cm]{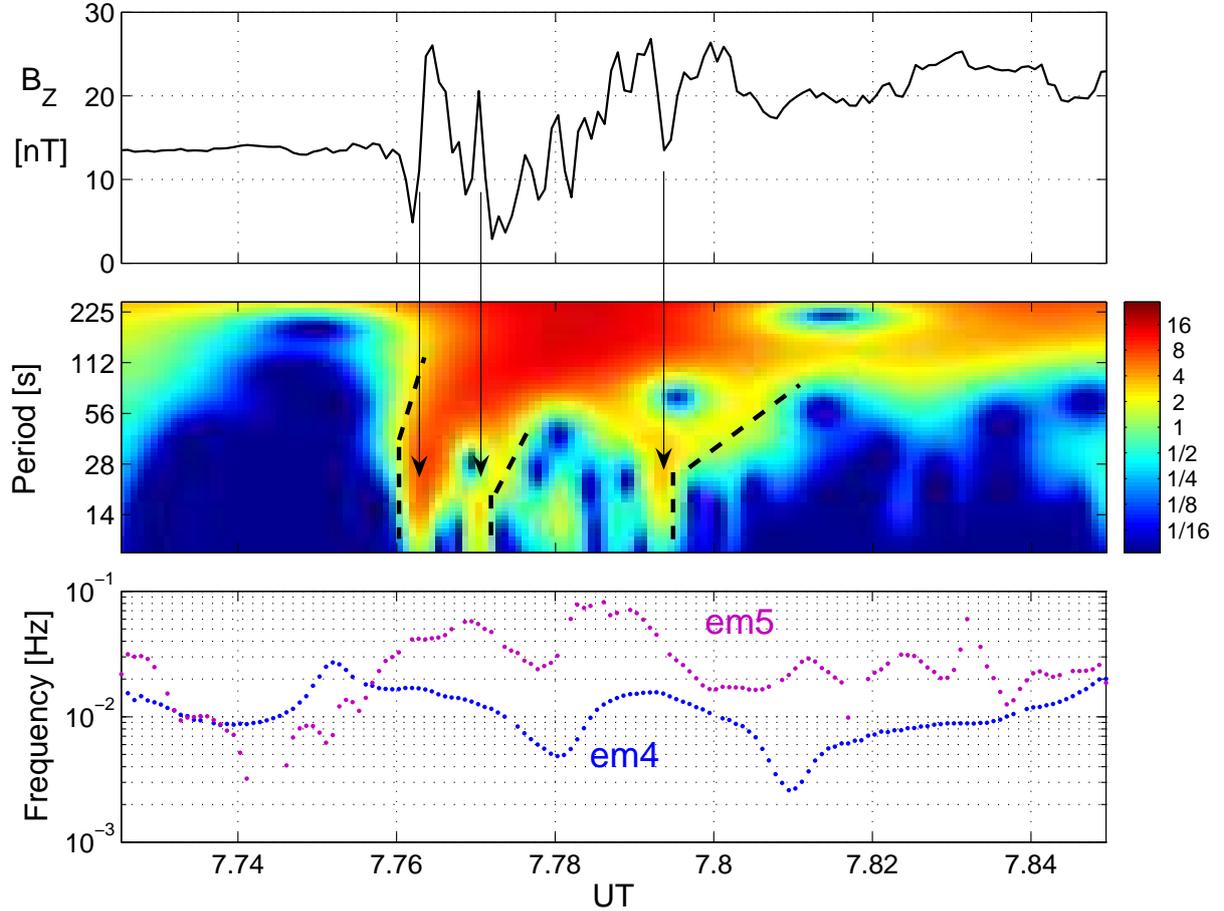}
 \caption{Top: magnetic field time series; Middle: Wavelet time-period-energy representation; Bottom: Hilbert-Huang
 time-instantenous frequency representation; Thick dashed lines in the middle indicate the wavelet time-scale evolution in time; Line arrows
 show the locations of short time-scale (high-frequency) activations occurring simultaneously with the largest jumps in $B_Z$.}
\end{center}
 \end{figure*}

The instantenous frequency is determined by taking HT of each $emi(t)$. The time and frequency
localization is demonstrated for the test signal in Figure 1c. The test signal was decomposed into three $emi$s (not shown).
Contrary to wavelet results (Figure 1b) the localization is rather good, without spurious energy distributions
in the time-frequency distribution. Now, we are going to apply the Hilbert-Huang method to the event studied by
\citet{lui08}, proclaimed to be an example of inverse cascade observed by the wavelet method.

\section{Identification of an inverse cascade}
CD and the associated dipolarization was observed by the THEMIS A spacecraft at $X_{GSM}\sim -8 R_E$,
between 07:45 and 07:50 UT on 29 January 2008. (In what follows, we will use decimal hours).
Since the event occurred near a substorm expansion onset, CD is considered to be
a substorm triggering agent. The inverse cascade was found to be most remarkable in the magnetic
component $B_Z$ \citep{lui08}.

The top panel in Figure 2 shows the $B_Z$ component of the magnetic
field. Using the Hilbert-Huang method described above six $emi$s
were found which represent the decomposition of $B_Z$ into
'monocomponent' modes. The modes are shown in the subsequent
subplots in Figure 2. In Figure 3 the power spectral density (PSD)
is depicted, computed for $B_Z$ (thick red line, for a better
visibility shifted) and  $em1-6$. The monocomponent modes are
narrow-band fluctuations, having finite bandwidth in PSD. For
example, $em2$ exhibits the maximum power around the frequency $\sim
4.5 10^{-2}$ Hz. Nevertheless, Figure 3 represents only a rough
demonstration of monocomponent modes. The modes in PSD correspond to
a global description based on the Fourier transform. The
monocomponent modes identified through the Hilbert-Huang method
differ due to the localized constraints introduced in their
determination explained above in Section 3. In spite of the finite
bandwith the Hilbert-Huang modes can exhibit  non-stationarity
through frequency and amplitude modulation \citep{hua98}.

Following the \citet{lui08} findings the inverse cascade should
operate roughly over the frequency range from $10^{-1}$ to
$5.10^{-3}$ Hz. The magnetic field PSD shows scaling (thick red
dashed line) over a narrower range from $3.10^{-2}$ to $5.10^{-3}$
Hz. Though we do not  intend to identify the inverse cascade on the
basis of PSD, this scaling region could be associated with the
developing nonlinear interactions. The power of $em1-3$ modes over
the scaling region is by 2-4 orders of magnitude smaller than the
power of $em4-6$ modes. Moreover, the $em6$ mode contains only a
smooth trend, therefore, we will investigate $em4-5$ further.

Figure 4 compares the wavelet time-period-energy spectrum (middle subplot) and the Hilbert-Huang time-frequency spectrum (bottom subplot).
$B_Z$ is in the top subplot. There are three vertical and inclined dashed lines in the wavelet plot. The lines are
guides for eyes to see the frequency trend of activity in time. At least in three cases, there is an increased wavelet power
which appears first at shorter time scales near the ion gyroperiod ($\sim$ 10 s). Power over longer time scales appears later.
Our wavelet representation of $B_Z$ is slightly different than the one published by \citet{lui08}.
Nevertheless, the wavelet representation shows the same qualitative features as
the wavelet analysis in \citet{lui08} paper: increasing wavelet power bridging gradually the scales between the ion
gyroperiod ($\sim$ 10s) and typical dipolarization time ($>$ 100 s).
It is easy to notice that the enhanced small-scale wavelet power
(indicated by dashed vertical lines in Figure 4) and the largest
amplitude jumps in $B_Z$ occur simultaneously (vertical arrow
lines). The dubious small-scale activity can occur due to a local
sharp change (like in Figure 1), which might have nothing to do with
the inverse cascade. Having in mind the limitations of the wavelet
method, the uncertainty connected with sharp jumps in the data
cannot be removed. Therefore, a wavelet independent test is needed.
The bottom subplot in Figure 4, corresponding to the Hilbert-Huang
time-frequency spectrum of $em4$ and $em5$ empirical modes (these
are within the PSD scaling region, see Figure 3), seems to
substantiate the idea of multi-scale physics. We note, the $emi$
spectra cannot be interpreted roughly before 07.74 and after 07.84
UT because of interval finite-size effects. The first high-frequency
activation (at $\sim 3.10^{-2}$ Hz after 07.75 UT) in $em4$ occurs
simultaneously with the beginning of small amplitude fluctuations in
$B_Z$. There is no increased power visible in the wavelet spectrum
at around 07.75 UT in Figure 4, neither in the wavelet spectrum of
\citet{lui08}, which could be associated with those $B_Z$
fluctuations. In $em4$ the frequency decreases to $5.10^{-3}$ Hz at
07.78 UT then, after a short return to $\sim 10^{-2}$ Hz, it falls
down to $\sim 3.10^{-3}$ Hz at 07.81 UT. Overall, the instantenous
$em4$ time scales change from 30 to 330 s as time proceeds. During
the interval of interest, $em5$ varies between $8.10^{-2}$ and
$10^{-2}$ (10-100 s). There are a few correlations between the
wavelet and Hilbert-Huang spectra in Figure 4. For example, the
enhanced small-scale activity in wavelet spectra is associated with
high-frequency fluctuations in $em5$ between 07.76 and 07.8 UT. The
first two local peaks in $em5$ seem to correlate with the first two
small-scale activations in the wavelet spectra between 07.76 and
07.77 UT. On the other hand, the local peak in $em5$ between 07.78
and 07.79 UT has no counterpart in the wavelet spectra. Moreover,
there are low-frequency valleys in between high-frequency local
peaks, which are related to the occurrence of multi-scale
interactions more complicated than a simple inverse cascade. In
contrary to the wavelet method the Hilbert-Huang results indicate
that the inverse cascade picture cannot fully explain the observed
multi-scale fluctuations. This does not necessarily mean that the
linkage between local kinetic instabilities and large-scale
processes is negligible.

\conclusions In this paper we have shown that the typical
limitations of wavelet methodology make the identification of CD
associated inverse cascades doubtful and misleading. The
observations of CD associated fluctuations are short in time, the
magnetic field changes are highly dynamic. As it was outlined in the
paper, the Hilbert-Huang method is more suitable for nonstationary
and nonlinear processes. Its adaptiveness allows to compute
straightforwardly the instantenous frequency and reproduce the
time-frequency representation of the time series based on decomposed
empirical modes. In qualitative agreement with the wavelet results,
the time evolution of instantenous frequency shows a general trend
of raising magnetic fluctuations from high-frequencies to
low-frequencies. However, the details of multi-scale fluctuations
and the time locations of high-frequency activations are rather
different in the wavelet and the Hilbert-Huang spectra. In the
latter, there are also intervals where the frequency locally
increases with time. In this paper we put a particular emphasis on
the differences between the wavelet and Hilbert-Huang
representations, demonstrating that for the CD associated
non-stationary magnetic fluctuations the wavelet approach can lead
to misleading conclusions.   We believe the more adaptive
Hilbert-Huang approach can reconstruct the time evolution of
instantenous frequency better, therefore, the identification of
turbulent cascades occurring over the appropriate frequency ranges
could be more straightforward. Since we were not able to identify a
simple inverse cascade from the data, the contribution of other
physical processes to the observed fluctuations cannot be ruled out.
For example, large-scale variations in the curvature of the ambient
magnetic field lines driven by ballooning instability prior to
substorm associated dipolarization onsets can lead to fluctuations
over the frequency range $10^{-1} - 10^{-2}$ Hz \citep{sait08}. This
is also the frequency range over which the inverse cascade proposed
by \citet{lui08} should occur. Global instabilities, including
magnetosphere-ionosphere coupling \citep{kan07} can also drive
low-frequency waves. As a matter of fact, the identification of
low-frequency fluctuations is rather difficult or impossible when
the whole length of event observations is comparable to the period
of waves. It is even more difficult to detect the non-stationarity
of low-frequency components over short intervals using wavelets. In
this paper it is shown that the low-frequency components of
fluctuations ($em4-5$ modes in Figures 3 and 4) are non-stationary
during the current disruption event of \citet{lui08}. As it was
mentioned in Section 2 non-stationarity of large-scale low-frequency
waves manifests itself over high-frequencies in the wavelet
spectrum, therefore the wavelet observations cannot be fully
trusted.

There exist also physical reasons which makes the identification of
turbulent cascades difficult. Inverse or forward cascades in
turbulence are related to local mode interactions in the Fourier
space. It is known that in the presence of a uniform magnetic field
magnetohydrodynamic turbulence is neither isotropic nor local in
Fourier space \citep{alex07}. There exist experimental indications
that such nonlocal interactions can occur near plasma boundaries
with developed gradients \citep{voro07}. The near-Earth location
where CD and magnetic field dipolarization occurs represents a
dynamically changing boundary between dipolar and more stretched
field lines. Spatial and temporal changes can be mixed up in
one-spacecraft in-situ observations and the multi-scale signatures
indicating the occurrence of an inverse cascade in the frequency
domain misleading.

\begin{acknowledgements}
We acknowledge NASA contract NAS5-02099 and V. Angelopoulos for use
of data from the THEMIS Mission. We thank K. H. Glassmeier and U.
Auster for the use of FGM data provided under the lead of the
Technical University of Braunschweig and with financial support
through the German Ministry for Economy and Technology and the
German Center for Aviation and Space (DLR) under contract 50 OC
0302. The work of Z.V. and M.P.L was supported by the Austrian
"Fonds zur F\"{o}rderung der wissenschaftlichen Forschung"  under
project P20131-N16.

\end{acknowledgements}

\end{document}